\documentclass[conference]{IEEEtran}

\pdfoutput=1

\usepackage{etoolbox}
\newbool{isgitdraft}
\boolfalse{isgitdraft}

\usepackage{xcolor}
\usepackage[normalem]{ulem}
\newbool{isReviewUpdate}
\boolfalse{isReviewUpdate}

\usepackage{cite}

\usepackage{multirow}

\usepackage[pdftex]{graphicx}
\graphicspath{{imgs/}}
\DeclareGraphicsExtensions{.pdf,.jpeg,.png}

\usepackage{amsmath}
\usepackage{amsfonts}
\usepackage{amsthm}
\usepackage{amssymb}
\usepackage{mathtools}

\usepackage{algpseudocode}
\usepackage{algorithm}

\algnewcommand{\LeftComment}[1]{\Statex \(\triangleright\) #1}

\usepackage{array}

\usepackage[caption=false,font=footnotesize]{subfig}

\usepackage{url}

\ifbool{isgitdraft}{
  \usepackage[mark]{gitinfo2}
}

\usepackage{./rfs_acros}

\usepackage{./rfs_macros}

\DeclareMathOperator*{\argmax}{argmax} % thin space, limits underneath in displays

\begin{document}

\ifbool{isgitdraft}{
  \begin{titlepage}
    \centering
    \vspace*{5cm}
    \huge\textit{\textbf{DRAFT}}: Deterministic Multi-sensor Measurement-adaptive Birth using Labeled Random Finite Sets \\
    \vspace{2\baselineskip}
    \large Compiled on \today\\
    \vspace{2\baselineskip}
    \textbf{Git Hash}: \texttt{\gitHash}\\
    \vspace{2\baselineskip}
    \textbf{Git Branch}: \texttt{\gitBranch}\\
    \vfill
  \end{titlepage}
  \newpage
}

\title{Deterministic Multi-sensor Measurement-adaptive Birth using Labeled Random Finite Sets}

\author{
  \IEEEauthorblockN{
    Jennifer Bondarchuk,
    Anthony~Trezza,
    Donald~J.~Bucci~Jr.,
  }
  \IEEEauthorblockA{
    \textit{Advanced Technology Laboratories} \\
    \textit{Lockheed Martin} \\
    Cherry Hill, NJ, USA \\
    \{jennifer.a.bondarchuk, anthony.t.trezza, donald.j.bucci.jr\}@lmco.com
  }
}

\maketitle
\begin{abstract}
    Measurement-adaptive track initiation remains a critical design requirement of many practical multi-target tracking systems.
    For labeled random finite sets multi-object filters, prior work has been established to construct a labeled multi-object birth density using measurements from multiple sensors.
    A truncation procedure has also been provided that leverages a stochastic Gibbs sampler to truncate the birth density for scalability.
    In this work, we introduce a deterministic herded Gibbs sampling truncation solution for efficient multi-sensor adaptive track initialization.
    Removing the stochastic behavior of the track initialization procedure without impacting average tracking performance enables a more robust tracking solution more suitable for safety-critical applications.
    Simulation results for linear sensing scenarios are provided to verify performance.
\end{abstract}


\IEEEpeerreviewmaketitle
% ================================================== %
% Introduction
% ================================================== %
\section{Introduction}\label{sec::intro}

\ac{MTT} is a sequential estimation problem where the goal is to jointly determine the number of objects and their trajectories from noisy, discrete-time measurements.
It is distinct from state estimation filtering in the sense that object appearance and disappearance (i.e., birth and death) must be handled alongside object detection and miss-detection in superimposed false measurements (i.e., clutter).
Approaches such as \ac{GNN} techniques~\cite{Blackman1999},~\ac{JPDA}~\cite{Barshalom2009}, and \ac{MHT}~\cite{Blackman2004} traditionally handle object appearance and disappearance as an implementation detail.
Labeled \ac{RFS}~\cite{Vo2013} has recently emerged as a family of techniques that include track birth and death in the problem formulation which is then solved via unified Bayesian framework.
We direct the reader to~\cite{Vo2015} for a detailed survey of the field, recent advances, and example applications.

The concept of conjugacy is of critical importance in labeled \ac{RFS} for maintaining computationally tractable \ac{MTT}~\cite{Vo2013}.
In the seminal works of~\cite{Vo2013,Vo2014,Vo2019}, it was shown that the \ac{GLMB} \ac{RFS} forms a conjugate multi-object prior with itself under the labeled \ac{RFS} Bayesian filtering equations as introduced in \ac{FISST} \cite{Mahler2007, Mahler2014}.
Under this construction the prior \ac{RFS} density for appearing targets is equivalently a labeled Poisson or a \ac{LMB} \ac{RFS}.
In a \emph{static birth} approach the prior \ac{RFS} density is assumed to be time invariant and established a priori.
In contrast, an \emph{adaptive birth} approach does not assume a time invariant birth distribution and must be constructed adaptively from the measurements \cite{Ristic2012}.

Measurement-adaptive birth techniques in \ac{RFS} began with the \ac{SMC} \ac{PHD} and \ac{CPHD} filters in~\cite{Ristic2012}, and later extended to their \ac{GM} variants in~\cite{Beard2013}.
Single-sensor adaptive birth techniques for labeled \ac{RFS} were introduced in~\cite{Reuter2014} for the~\ac{LMB} filter and~\cite{Lin2016} for the~$\delta$-\ac{GLMB} filter.
These approaches follow the concepts presented in~\cite{Ristic2012} by constructing a component of an \ac{LMB} RFS from every single-sensor measurement observed at the previous time step.
The probabilities of birth for each of these \ac{LMB} components were set proportionally to a maximum birth rate parameter.
The proportionality factor was based on how much a measurement did not associate with the persisting targets from the last time step.
In contrast, the multi-sensor adaptive birth regime carries the additional complication that there are exponentially many possible measurement tuples that could be used to generate newborn targets.
In our previous work~\cite{Trezza2022}, we proposed a stochastic Gibbs sampling approach for truncating multi-sensor measurement tuples that would likely be pruned in the subsequent $\delta$-\ac{GLMB} update step.
The proposed Gibbs sampler exhibited quadratic complexity in the number of sensors, and both an \ac{SMC} approximation and a closed form Gaussian solution were provided.

The drawback of using a stochastic Gibbs sampler for multi-sensor adaptive birth is that it adds non-determinism during track initiation.
Stochasticity during track initiation can cause substantial differences in the resulting track output, which can negatively impact tracking performance.
This run-to-run variability hinders verification and validation testing in production systems, especially when safety-guaranteed operation is critical \cite{Koopman2016}.

In this work, we draw inspiration from~\cite{Wolf2020} to construct a deterministic Gibbs sampler for the multi-sensor adaptive birth technique presented in~\cite{Trezza2022}.
The proposed deterministic, multi-sensor adaptive birth procedure leverages a deterministic sampling technique known as \emph{herding}~\cite{Welling2009} as applied to Gibbs sampling~\cite{Chen2016}.
As we will show, this deterministic multi-sensor adaptive birth procedure converges faster, requiring fewer samples, and maintains the average tracking performance presented in~\cite{Trezza2022}.
It has the additional benefit that for the closed-form Gaussian variation, the proposed approach is fully deterministic.

The remainder of this paper is organized as follows.
Section~\ref{sec::background} provides background material on the relevant labeled~\ac{RFS} terminology and a brief review of the multi-sensor adaptive birth technique of~\cite{Trezza2022}.
Section~\ref{sec::herdedabi} provides a brief background on deterministic sampling via herding and introduces the deterministic multi-sensor adaptive birth procedure.
Finally Section~\ref{sec::sim} compares the proposed deterministic adaptive birth technique against the stochastic version of~\cite{Trezza2022} within the~\ac{LMB} and~$\delta$-\ac{GLMB} filters.

% ================================================== %
% Background
% ================================================== %
\section{Background}\label{sec::background}
% ================================================== %
% Notation
% ================================================== %

\subsection{Notation}
We adopt the notation of~\cite{Vo2013, Vo2019, Trezza2022} as follows.
Lowercase letters denote vectors (e.g., $x, \textbf{x}$), whereas uppercase letters denote finite sets (e.g., $X, \textbf{X}$).
Bold letters denote labeled states and their distributions (e.g., $\textbf{x}, \textbf{X}, \boldsymbol{\pi}$) and blackboard bold letters denote spaces (e.g., $\mathbb{X}, \mathbb{Z}, \mathbb{L}, \mathbb{R}$).
Variable sequences $X_i, X_{i+1}, \dots, X_j$ are abbreviated $X_{i:j}$.
The inner product $\int f(x) g(x) dx$ is written as $\langle f, g \rangle$.
Finally, we drop the subscript notation for current time step $k$ and use subscript '$+$' to indicate the next time step ($k+1$).

% ================================================== %
% MSJAB
% ================================================== %
\subsection{Multi-sensor Multi-target Adaptive Birth Model}
Our objective is to adaptively construct an \ac{LMB}~\ac{RFS} density $\textbf{f}_{B,+}$ given the multi-object measurement sets $Z^{(s)} \in \mathbb{Z}^{(s)}$ from sensors $s \in \{1, \dots, V\}$.
Let $m^{(s)} = |Z^{(s)}|$ and $\mathbb{J}^{(s)} \triangleq \{1, \dots, m^{(s)}\}$ be an index set into $Z^{(s)}$.
Then let $\mathbb{J}^{(s)}_0 \triangleq \mathbb{J}^{(s)} \cup \{0\}$, where the entry $0$ represents a miss-detection.
The tuple $J \triangleq (j^{(1)}, \dots, j^{(V)}) \in \mathbb{J}^{(1)}_0 \times \dots \times \mathbb{J}^{(V)}_0$ is a multi-sensor measurement index with $Z_J \triangleq (z^{(1)}_{j^{(1)}}, \dots, z^{(V)}_{j^{(V)}})$ as the corresponding tuple of multi-sensor measurements.

In the single-sensor multi-target observation model of \cite{Vo2013, Vo2014}, a target state $\textbf{x} \in \textbf{X}$ is either detected by sensor $s$ with probability $p_D^{(s)}(\textbf{x})$ or miss-detected with probability $1 - p_D^{(s)}(\textbf{x})$.
Target measurements $z^{(s)}$ are generated via the measurement likelihood function $g^{(s)}(z^{(s)}|\textbf{x})$.
The detection dynamics of a target are modeled as a multi-Bernoulli \ac{RFS} with parameter set $\{(p_D^{(s)}(\textbf{x}), g^{(s)}(z^{(s)}|\textbf{x})): \textbf{x} \in \textbf{X}\}$, with conditional independence between each multi-Bernoulli RFS given $\textbf{X}$.
Each sensor generates clutter detections according to a Poisson RFS with intensity function $\kappa^{(s)}$.
The superposition of the target and clutter detections form the multi-object observation $Z^{(s)}$.
The multi-object measurement likelihood is the convolution of the target detection and clutter detection RFS distributions.
The exact form of this likelihood is omitted for brevity here, and can be found in \cite{Vo2013, Vo2015}.
An important quantity within the multi-object likelihood relevant to this work is the per-object, per-sensor measurement \emph{pseudolikelihood}
\begin{equation}\label{eq::meas_likelihood}
	\psi^{s, j^{(s)}}_{Z^{(s)}}(\textbf{x}) =
	\begin{cases}
		\frac{p_D^{(s)}(\textbf{x})g^{(s)}(z^{(s)}_{j^{(s)}} | \textbf{x})}{\kappa^{(s)}(z^{(s)}_{j^{(s)}})} & j^{(s)} \in \mathbb{J}^{(s)} \\
		1 - p_D^{(s)}(\textbf{x})                                                                            & j^{(s)} = 0
	\end{cases}.
\end{equation}

As suggested in~\cite{Trezza2022}, the birth \ac{LMB}~\ac{RFS} density used at time ($k+1$) is modeled using the measurements at time $k$ as
\begin{equation}
	\textbf{f}_{B,+} = \left\{ \left(r_{B,+}(l_+), p_{B,+}(\cdot, l_+ | Z_J) \right) \right\}_{l_+ \in \mathbb{B}_+},
\end{equation}
where $l_+ \in \mathbb{B}_+ = \{(k+1, J) : \forall J \in \mathbb{J}_0\}$ is the newborn object's label having birth probability $r_{B,+}(l_+)$.
The quantity $p_{B,+}(\cdot, l_+ | Z_J)$ is the state distribution of label $l_+$ conditioned on the measurement tuple $Z_J$.
The state distribution is given by the Chapman-Kolmogrov equation \cite{Mahler2007}
\begin{equation}\label{eq::pred_birth_post}
	p_{B,+}(x_+, l_+ | Z_J) = \int f_+(x_+|x, l) p_B(x, l_+ | Z_J) dx,
\end{equation}
where,
\begin{align}
	p_B(x, l_+ | Z_J)       & = \frac{p_B(x, l_+) \psi^{J}_Z(x, l_+)}{\bar{\psi}^{J}_{Z}(l_+)} \label{eq::spatial_distr} \\
	\bar{\psi}^{J}_{Z}(l_+) & = \langle p_B(\cdot, l_+), \psi^{J}_{Z}(\cdot, l_+)\rangle. \label{eq::psi_bar}
\end{align}
Here $f_+(x_+|x, l)$ is the Markov transition density modeling the dynamics of the object state and $p_B(x, l_+)$ is the prior distribution on the newborn object's state.
The quantity $\psi^{J}_{Z}(x, l_+)$ is the \emph{multi-sensor measurement pseudolikelihood} function \cite{Vo2019}.
Assuming conditional independence of the sensor measurements,
\begin{equation}\label{eq::joint_likelihood}
	\psi^J_Z(\textbf{x}) \triangleq \prod\limits^V_{s=1} \psi^{s, j^{(s)}}_{Z^{(s)}}(\textbf{x}).
\end{equation}

The birth probability of label $l_+$ is modeled in \cite{Trezza2022} as
\begin{equation}\label{eq::birth_prob}
	r_{B,+}(l_+) = \min\left(r_{B, \text{max}}, \hat{r}_{B,+}(l_+)\lambda_{B,+} \right),
\end{equation}
where $r_{B, \text{max}} \in [0, 1]$ is the maximum existence probability of a newborn target and $\lambda_{B,+}$ is the expected number of target births at time step $k+1$.
The effective birth probability $\hat{r}_{B,+}(l_+)$ is given as
\begin{equation}\label{eq::birth_prob_hat}
	\hat{r}_{B,+}(l_+) = \frac{r_U(J) \bar{\psi}^{J}_{Z}(l_+)}{\sum\limits_{J' \in \mathbb{J}_0} r_U(J')\bar{\psi}^{J'}_{Z}(l_+)},
\end{equation}
where $r_U(J)$ is the probability that the multi-sensor measurement tuple $J$ does not associate with any existing targets.
This non-association probability is approximated as
\begin{equation}\label{eq::unassoc_prob}
    r_{U}(J) \propto \prod_{j \in J} \left(1 - r_{A}(j)\right),
\end{equation}
where $r_{A}(j)$ is the probability that a single-sensor's measurement, $z^{(s)}_{j^{(s)}}$, is associated with existing targets in the $\delta$-\ac{GLMB} posterior density.
For more information on these quantities, we direct the reader to~\cite{Trezza2022}.

% ================================================== %
% Gibbs Sampling Truncation
% ================================================== %
\subsection{Truncating the Multi-sensor Adaptive Birth LMB}
The number of newborn labels $\mathbb{B}_+$ is driven by the cardinality of $\mathbb{J}_0$ which has $O(m^V)$ elements, where $m$ is the worst case number of measurements.
It is therefore not practical to enumerate every possible entry at each time step.
In the spirit of \cite{Vo2017}, a technique was proposed in \cite{Trezza2022} to sample labels $\mathbb{B}'_+ \subset \mathbb{B}_+$ that have high $\hat{r}_{B,+}(l_+) $ and hence are likely to persist through future measurement updates.
The sampling distribution is modeled as
\begin{equation}\label{eq::orig_sampling_distr}
	p(l_+) \propto \hat{r}_{B,+}(l_+) \propto r_U(J)\bar{\psi}^{J}_{Z}(l_+).
\end{equation}
Directly sampling from the categorical distribution $p(l_+)$ requires evaluation of an exponential number of possible birth labels.
By construction of the birth set, sampling from $p(l_+)$ is equivalent to sampling multi-sensor measurement tuples from the joint distribution, $p(J) = p(j^{(1)}, \dots, j^{(V)})$.
A Gibbs sampler was thus proposed in \cite{Trezza2022} to achieve efficient sampling from Equation~(\ref{eq::orig_sampling_distr}) using the Markov transition probabilities
\begin{equation}\label{eq::cdn_likelihood}
	p(j^{(s)} | J^{-s}) \propto \left(1 - r_A(j^{(s)})\right) \bar{\psi}^{J}_{Z}(l_+),
\end{equation}
where $J^{-s} = (j^{(1)}, \dots, j^{(s-1)}, j^{(s+1)}, \dots, j^{(V)})$.

The stochastic Gibbs sampler that generates the measurement tuples (i.e., birth components) with significant birth probabilities is given in \cite[Algorithm 1]{Trezza2022}.
Similar to \cite{Vo2017}, it exhibits an exponential convergence rate and does not require a burn-in period since unique solutions can be directly used in construction of the birth set.
An \ac{SMC} method was provided in \cite[Section VI]{Trezza2022} for evaluating the transition probabilities and generating the state distributions.
Closed-form solutions for these quantities in the Gaussian case can be found in \cite[Section VII]{Trezza2022}.

% ================================================== %
% Adaptive Birth with Herded Gibbs Sampling
% ================================================== %
\section{Herded Gibbs Multi-sensor Adaptive Birth}\label{sec::herdedabi}

\subsection{Deterministic Sampling via Herding}

%%%%%%%%%%%%%%%%%%
As discussed in \cite{Welling2009, Wolf2020}, herding is a deterministic procedure
to generate a sequence of pseudo-samples that match the moments, $\mu = \mathbb{E}_{x \sim P}\left\{ \phi(x) \right\}$, with respect to the features, $\phi(x)$, where $P$ is a discrete probability distribution.
It works using a collection of nonlinear update equations to solve a minmax problem.
The first phase selects a state such that the objective function is minimized.
The second phase updates auxiliary weights to maximize the objective function.
The $i^{\text{th}}$ sample $\gamma^{(i)} \in \mathbb{G}$ is computed as
\begin{align}
    \gamma^{(i)} =  \argmax_{\gamma \in \mathbb{G}}  \langle w^{(i - 1)}, \phi(\gamma) \rangle \label{eq::herding_update1}\\
    w^{(i)} = w^{(i-1)} + \mu - \phi(\gamma^{(i)}) \label{eq::herding_update2},
\end{align}
where $w \in \mathbb{W}$ is a weight vector that includes information on all previous samples, and $\phi : \mathbb{G} \rightarrow \mathbb{W}$ is a feature map.
If $\phi$ is an indicator function, it enables sampling from a probability distribution with $\mu = P$.
From here forward we will drop the superscript notation for the $i^{\text{th}}$ iteration for brevity and to avoid confusion with our superscript indexing notation.

%%%%%%%%%%%%%%%%%%

In \cite{Chen2016}, the herding procedure was applied to the discrete conditional distribution in a Gibbs sampler.
A set of all weight vectors $W$ is maintained containing the weight vector for all Markov transitions distributions.
For space efficiency, this can be maintained in a dictionary mapping $W: \mathbb{T} \times \mathbb{N} \rightarrow \mathbb{W}$,
such that $W(\gamma, n) \equiv w_{(\gamma, n)}$ is the weight vector for the transition distribution for the $n^{\text{th}}$ state of vector $\gamma$.
The dictionary $W$ is initialized as empty and $w_{(\gamma, n)}$ is initialized proportional to the transition density.
On every iteration, samples are generated by computing the argmax of $w_{(\gamma, n)}$ per Equation~(\ref{eq::herding_update1}) and $w_{(\gamma, n)}$ is updated according to Equation~(\ref{eq::herding_update2}).

\subsection{Deterministic Multi-sensor Adaptive Birth}
The herded Gibbs sampler for the multi-sensor adaptive birth truncation procedure is provided in Algorithm~\ref{alg::gibbs}.
In practice, $p(j^{(s)} | J^{-s})$ can be computationally complex to compute for arbitrary transition densities since it requires the evaluation of multi-variate, indefinite integrals.
To alleviate the need to compute this repeatedly, we suggest caching the value in a dictionary mapping, $M : \mathbb{T} \times \mathbb{N} \rightarrow \mathbb{R}^V$ so it can be accessed in $O(1)$ on future iterations that revisit state $J$ and are transitioning for sensor $s$.

As discussed in \cite{Trezza2022}, tempering is required to ensure sufficient exploration such that all significant birth components are generated.
The need for tempering is especially apparent with herding Gibbs sampling, which has been observed to exhibit improved sample efficiency over its stochastic counterparts \cite{Chen2016, Wolf2020}.
To alleviate this, sensor order is lexicographically cycled at each Gibbs iteration, and is denoted by the $\text{Perm}$ function in Algorithm~\ref{alg::gibbs}.
Additionally, all values of $\bar{\psi}^{J}_{Z}(l_+)$ can be capped to avoid poorly scaled transition distributions and birth probabilities, while still allowing for components with low values to be pruned.

The asymptotic complexity of Algorithm~\ref{alg::gibbs} is driven by the implementation of $p(j^{(s)} | J^{-s})$ (see \cite{Trezza2022}).
In the worst-case, $p(j^{(s)} | J^{-s})$ needs to be evaluated the same number of times as the stochastic variant proposed in \cite{Trezza2022}.
The complexity of Algorithm~\ref{alg::gibbs} is equivalent to the complexity of Algorithm 1 in \cite{Trezza2022}.

\begin{algorithm}[t!]
    \caption{Adaptive Birth Herded Gibbs Sampler}\label{alg::gibbs}
        \begin{algorithmic}[1]
            \renewcommand{\algorithmicrequire}{\textbf{Input:}}
            \renewcommand{\algorithmicensure}{\textbf{Output:}}
            \Require
                \Statex $Z$, $r_U$, $p_{B,+}$, $T$, $k$, $V$
            \Ensure
                $\mathbb{B}'_+$
            \State $\mathbb{B}'_+ = \emptyset$
            \State $J = \left(0, \dots, 0\right)$
            \State $W = \{\}$, $M = \{\}$
            \For{$t = 1, \dots, T$}
                \For{$s \in \text{Perm}(\{1, \dots, V\})$}
                    \For{$j^{(s)} \in \mathbb{J}^{(s)}_0$}
                    \State \(\triangleright\) Query the prior weight vector from cache
                        \If{$(J, s) \notin W$}
                            \State $\mu_{(J,s)} = p(j^{(s)} | J^{-s})$ \cite[Theorem V.1]{Trezza2022}
                            \State $w_{(J,s)} = \mu_{(J,s)}$
                            \State $M(J, s) = \mu_{(J,s)}$
                        \Else
                            \State $w_{(J,s)} = W(J, s)$
                            \State $\mu_{(J,s)} = M(J, s)$
                        \EndIf
                    \EndFor
                    \State \(\triangleright\) Find the next sample in the Gibbs chain
                    \State $j'^{(s)} = \argmax(w_{(J,s)})$
                    \State $J' = (j^{(1)}, \dots, j^{(s-1)}, j'^{(s)}, j^{(s+1)}, \dots, j^{(V)})$
                    \State $\mathbb{B}'_+ = \mathbb{B}'_+ \cup (k, J')$
                    \State \(\triangleright\) Performing the herding update on the weights
                    \State $w_{(J,s)} = w_{(J,s)} + \mu$
                    \State $w_{(J,s)}^{(s)} = w^{(s)}_{(J,s)} - 1$
                    \State $W(J, s) = w_{(J,s)}$

                \EndFor

                \EndFor
        \end{algorithmic}
    \end{algorithm}

% ============================================================== %
% Simulations
% ============================================================== %
\section{Simulations}\label{sec::sim}
\subsection{Overview}\label{sec::overview}
In this section we provide results from two simulations to demonstrate the performance of the proposed herded Gibbs adaptive birth sampler.
In the first simulation, the herded Gibbs adaptive birth sampler was compared to the stochastic variant in both the Gaussian mixture $\delta$-\ac{GLMB} and \ac{LMB} filters.
This simulation used the stochastic update ranked assignment procedure from \cite{Vo2017}, which is used to produce simulation results in \cite{Trezza2022}.
This verifies the performance and convergence rate of the proposed herded Gibbs sampler.
In the second simulation, we compare fully deterministic variants of the Gaussian mixture $\delta$-\ac{GLMB} and \ac{LMB} filters by using a herded Gibbs sampler in the update step.
The herded Gibbs adaptive birth sampler is used with the herded Gibbs update sampler \cite{Wolf2020} and the stochastic Gibbs update sampler \cite{Vo2017}.
This verifies the performance of a fully deterministic filter.

\begin{figure}[t!]
    \includegraphics[width=0.48\textwidth, height=6cm]{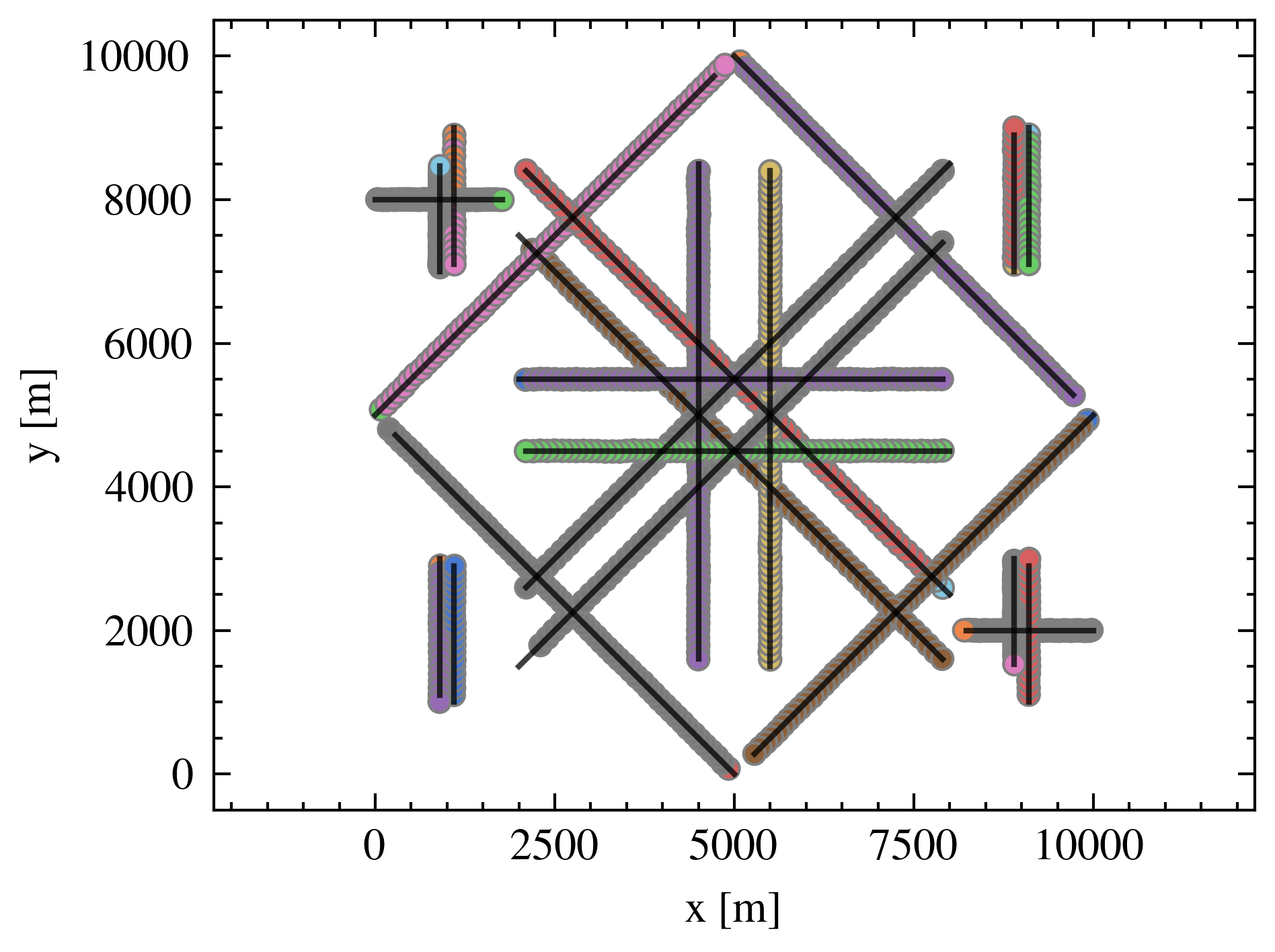}
    \caption{Single run results using deterministic multi-sensor adaptive birth model.
    Labeled target position estimates (colored circles, each color corresponds to a unique track label), target trajectories (black lines), target birth locations (black circles).
    }
    \label{fig::qual}
\end{figure}

The scenario tracks a time-varying number of targets in planar 2D position and velocity, $x = [p_{x}, \dot{p}_x, p_y, \dot{p}_y]^T$.
Target dynamics were simulated according to a constant velocity transition model, $f_+(x_+|x) = \mathcal{N}(x_+; Fx, Gw)$ where

\begin{equation*}
    F =
    \begin{bmatrix}
            1 & \Delta\\
            0 &     1
    \end{bmatrix},\qquad
    G =
    \begin{bmatrix}
        \frac{\Delta^2}{2}\\
        \Delta
    \end{bmatrix},
\end{equation*}
under the discrete-time sampling interval $\Delta$ \cite{Li2003}.
The driving noise $w$ was simulated as zero-mean Gaussian with $\textnormal{Cov}(w) = \textnormal{diag}[5, 5]^T$.
The survival probability for each target was $p_s(x,l) = 0.99$.
The target trajectories were generated according to Figure~\ref{fig::qual}, with at most 22 targets occurring at a single time step.
The birth locations of each target were fixed, but are sparse with no two targets being born from the same location.

Eight linear position sensors were simulated, each having detection probability $p_D^{(s)}(x,l) = 0.95$.
If an object was detected, the measurement $z^{(s)} = [p_x, p_y]^T$ was observed according to the single-target measurement likelihood $g(z^{(s)}|x) = \mathcal{N}(z^{(s)}; H^{(s)}x, R^{(s)})$ with $R^{(s)} = \text{diag}(10^2, 10^2)$ and, $H^{(s)} = \begin{bsmallmatrix}1 & 0\\0 & 1\end{bsmallmatrix} \otimes \begin{bsmallmatrix}1 & 0\\0 & 0\end{bsmallmatrix}$ where $\otimes$ denotes the Kronecker product.
Clutter was Poisson distributed with intensity $\kappa^{(s)}(\mathbb{Z}^{(s)}) = \lambda_c^{(s)} \mathcal{U}(\mathbb{Z}^{(s)})$ where $\mathcal{U}(\mathbb{Z}^{(s)})$ is the uniform distribution over $\mathbb{Z}^{(s)}$, and $\lambda^{(s)}_c = 15$.

Each filter was simulated over 100 Monte Carlo iterations.
The prior was modeled as a Gaussian with an uninformative covariance in position such that, $p_B(x, l_+) = \mathcal{N}(x; \mu_0, P_0)$ with $\mu_0 = \left[0, 0, 0, 0\right]^T$ and $P_0 = \text{diag}(100000^2, 50^2, 100000^2, 50^2)$.
Both adaptive birth Gibbs samplers were configured to use 250 iterations per evaluation, which is reduced from 1000 iterations as used in \cite{Trezza2022}.
The remaining adaptive birth parameters were kept the same.

%%%%%%%%%%%%%%%%%%%%%%%
\subsection{Results}\label{sec::sim::results}

\begin{figure*}[t!]
    \begin{minipage}[t]{0.48\textwidth}
    \subfloat[$\delta$-\ac{GLMB} filter]{\includegraphics[height=6cm]{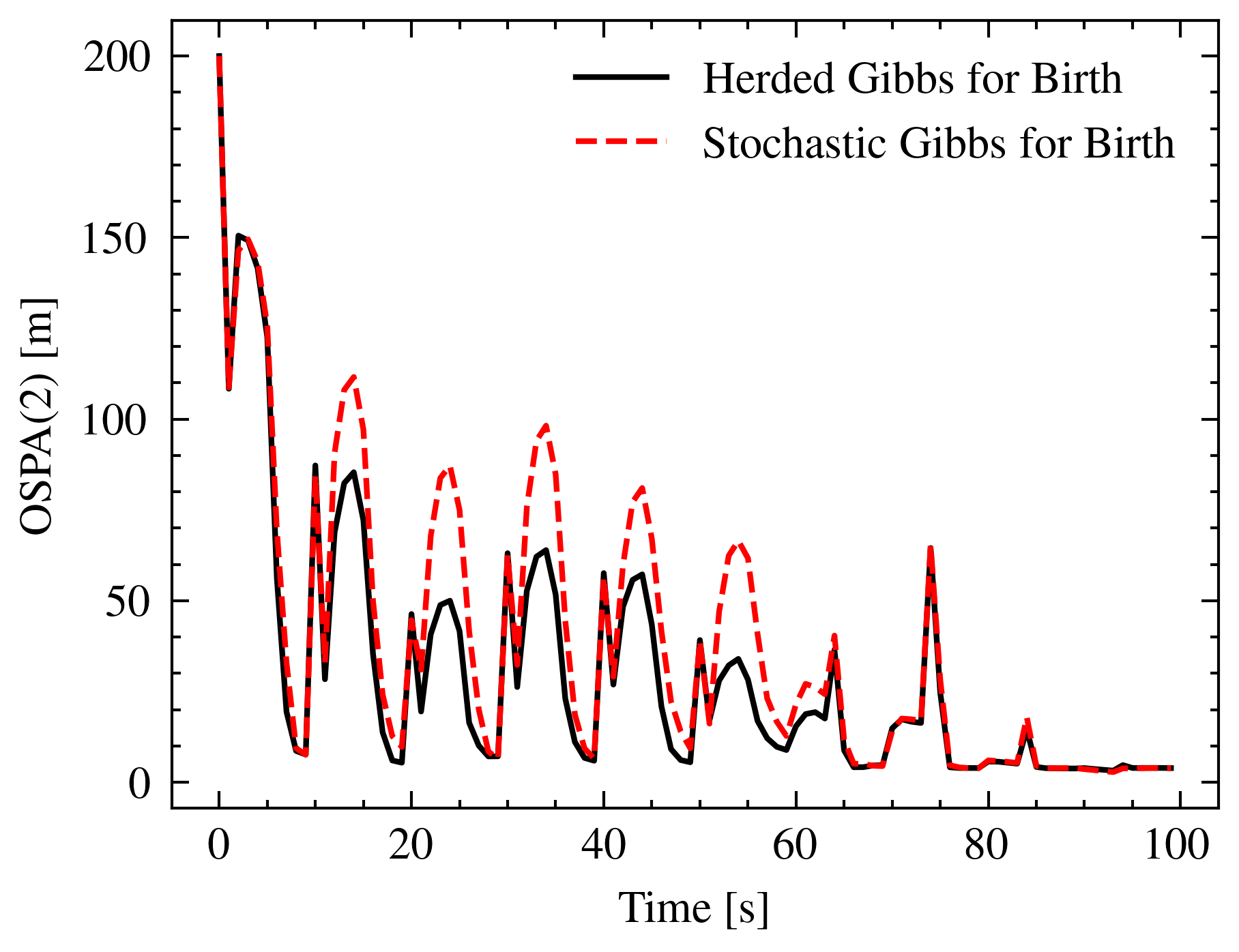}\label{fig::glmb::results::ospa}}
    \newline
    \subfloat[\ac{LMB} filter]{\includegraphics[height=6cm]{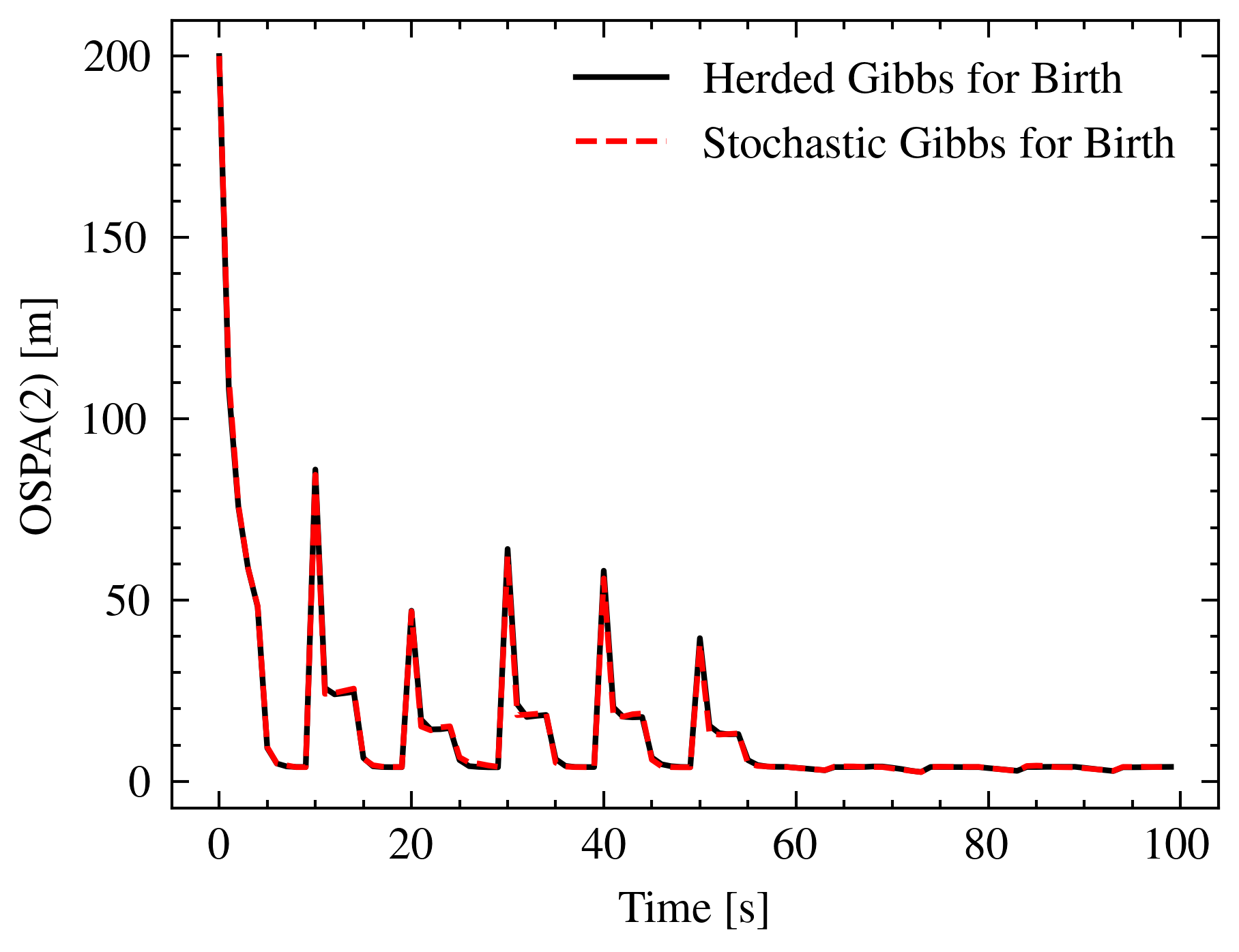}\label{fig::lmb::results::ospa}}
    \caption{\ac{OSPA}(2) results comparing herded and stochastic adaptive birth Gibbs samplers. These simulations were configured to use the stochastic Gibbs sampler for ranked assignment to match \cite[Section VIII.B.]{Trezza2022}.}
    \label{fig::results::ospa}
    \end{minipage}
    \hfill
    \begin{minipage}[t]{0.48\textwidth}
    \subfloat[$\delta$-\ac{GLMB} filter]{\includegraphics[height=6cm]{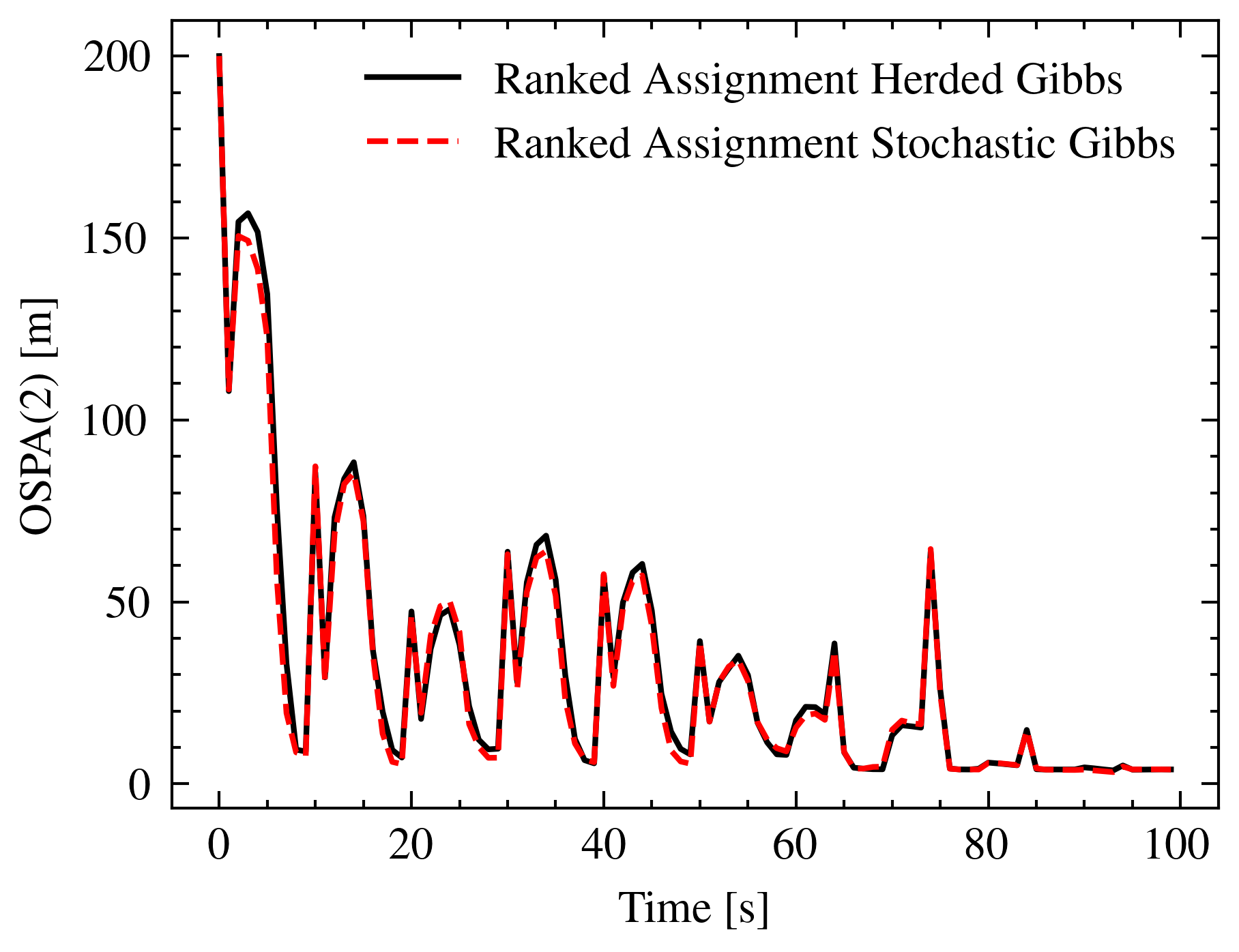}\label{fig::glmb::results::average_ospa_ranked_assignment}}
    \newline
    \subfloat[\ac{LMB} filter]{\includegraphics[height=6cm]{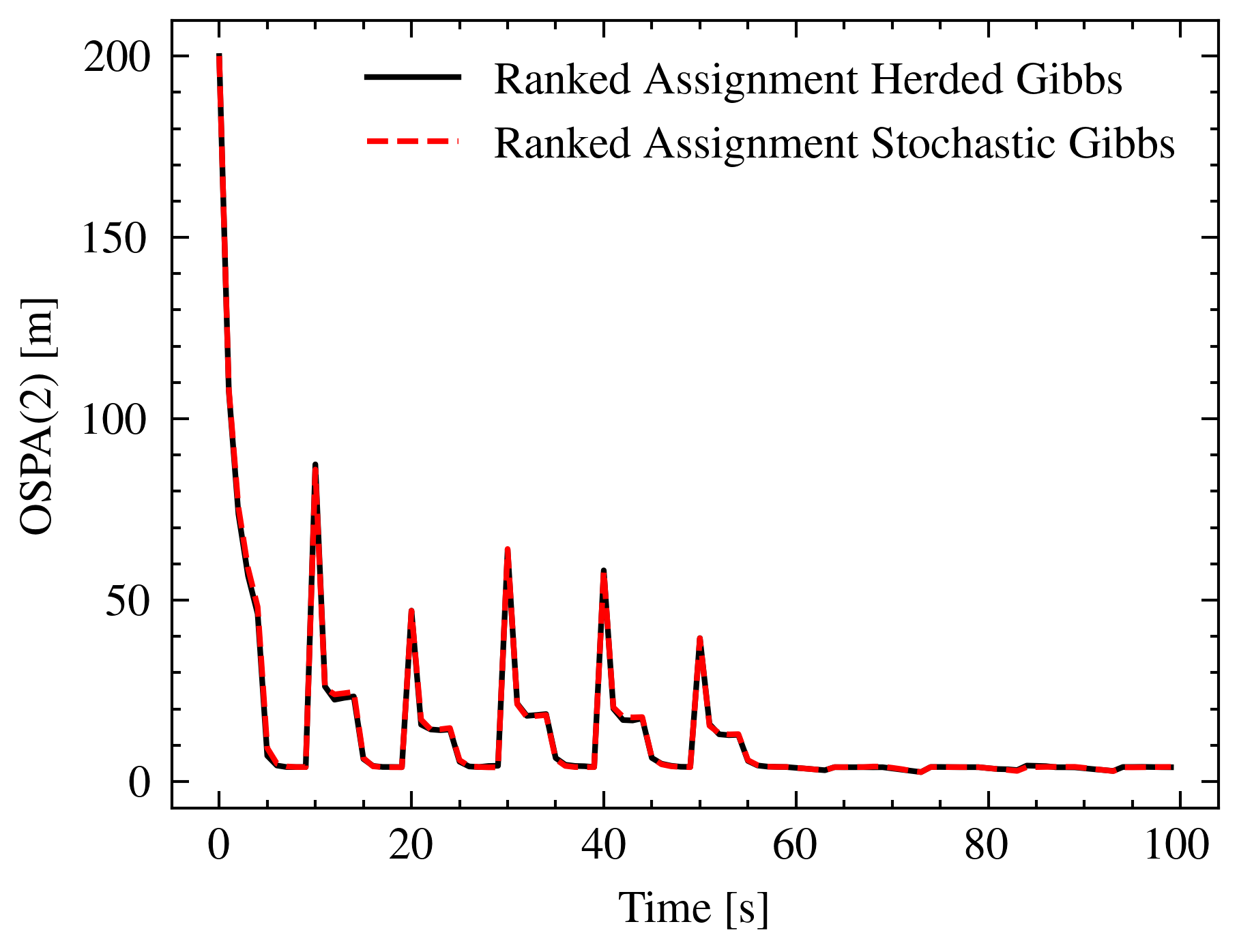}\label{fig::lmb::results::average_ospa_ranked_assignment}}
    \caption{\ac{OSPA}(2) results comparing herded and stochastic ranked assignment Gibbs samplers. These simulations were configured to use the herded Gibbs sampler for birth.}
    \label{fig::results::average_ospa_ranked_assignment}
    \end{minipage}
\end{figure*}

The cardinality and state estimation accuracy were estimated using the \ac{OSPA}(2) metric \cite{Beard2020} as depicted in Figure~\ref{fig::results::ospa} and Figure~\ref{fig::results::average_ospa_ranked_assignment}.
The \ac{OSPA}(2) metric was computed using a distance cutoff value of $200$, a distance order of $1.0$, a sliding window length of $5$ and an expanding window weight power of $0$.

As shown in Figure~\ref{fig::results::ospa}, the proposed herded Gibbs adaptive birth sampler had a lower average OSPA(2) score in the $\delta$-\ac{GLMB} filter simulations.
Both Gibbs samplers performed similarly for the \ac{LMB} filter.
These results show that the herded Gibbs sampler converged faster, using fewer iterations, and achieved similar performance to \cite{Trezza2022}.

The average OSPA(2) scores in Figure~\ref{fig::results::average_ospa_ranked_assignment} is similar for both the $\delta$-\ac{GLMB} and \ac{LMB} filters.
The best performance is achieved using the herded Gibbs sampler for birth, with either Gibbs sampler for ranked assignment.

% ============================================================== %
% Conclusions
% ============================================================== %
\section{Conclusions}\label{sec::conclusions}

This paper provided a deterministic Gibbs truncation algorithm for multi-sensor measurement-adaptive birth.
This removes stochasticity from a tracking system, reducing the run-to-run variability of a tracker's performance.
The main advantage of this approach is that it guarantees reproducible results, which makes it more suitable for use in safety critical operations.
The results of the simulations show the average tracking performance was similar to or better than the stochastic Gibbs sampler, and uses fewer iterations than\cite{Trezza2022}.
Future work includes analysis to find the minimum required iterations to match the stochastic ranked assignment Gibbs sampler performance.

\bibliographystyle{IEEEtran}
\bibliography{IEEEabrv, sections/ms.bib}

\end{document}